# Comprehensive indicator comparisons intelligible to non-experts: The case of two SNIP versions [1]

Henk F. Moed

Visiting professor, Sapienza University of Rome, Department of Computer, Control & Management Engineering A. Ruberti, Via Ariosto 25, I-00185 Rome, Italy. Email: henk.moed@uniroma1.it

**Abstract**

A framework is proposed for comparing different types of bibliometric indicators, introducing the notion of an Indicator Comparison Report. It provides a comprehensive overview of the main differences and similarities of indicators. The comparison shows both the strong points and the limitations of each of the indicators at stake, rather than over-promoting one indicator and ignoring the benefits of alternative constructs. It focuses on base notions, assumptions, and application contexts, which makes it more intelligible to non-experts. As an illustration, a comparison report is presented for the original and the modified SNIP (Source Normalized Impact per Paper) indicator of journal citation impact.

1. **Introduction**

More and more indicators are being proposed, some explicitly competing with an existing indicator, others rather complementary. There is an institutional pressure to launch new indicators, both within the academic environment, and within the domain of scientific information companies. Especially the past few years a lively debate has taken place about bibliometric indicators which focuses on their statistical properties. A series of articles have proposed new types of *journal* indicators, including González-Pereira, Guerrero-Bote &Moya-Anegón (2010) on the Scimago Journal Rank (SJR); Falagas, Kouranos, Arencibia-Jorge & Karageorgopoulos (2008) comparing SJR with classical journal impact factors; Bornmann, Marx & Schier (2009) on Hirsch-type of journal metrics; Fersht (2009) comparing Eigenfactor with journal impact factors; Harzing & Van der Wal (2009) on a Google Scholar-based Hirsch index for journals; Pislyakov (2009) comparing journal impact factors from Thomson Reuters' Journal Citation Reports with journal indices derived from Elsevier's Scopus; Yin, Aris & Chen (2010) combining Eigenfactor and Hirsch Index for journals; Leydesdorff and Opthof (2010) comparing SNIP with another indicator based on fractional counting; Campanario (2011) comparing journal impact factors based on two- and on five-year citation windows; Glänzel, Schubert, Thijs & Debackere (2011) on a posteriori and a priori citation-based indicators; Leydesdorff & Bornmann (2011) on journal metrics based on fractional counting; Glanzel & Moed providing an overview of major statistical issues in indicator development ;and Mingers (2014) comparing the original and modified SNIP.

The launch of so many new indicators creates the need to systematically compare indicators, not only for the members of the bibliometric research community, for whom systematic comparisons constitute a first step towards standardization of indicators, but also for the various groups of users of the indicators. If one indicator is in a certain respect better than another, or if two indicators show a strong correlation across their entire data range, users should be informed about this in a proper manner. In the first case they may consider to use a new indicator, while in the second they need not bother too much about which one they should use.

---

[1]This paper is based on a presentation given by the author at the Workshop on Bibliometric Indicators, held at Leiden University, and organized by the Center for Science and Technology Studies (CWTS) on September 2, 2014

The aim of the present paper is not to present a comprehensive overview and critical discussion of all types of citation impact indicators that have been explored during the past decade. It proposes a general framework for the comparison of different types of indicators, based on the notion that it is fair to highlight both the limitations and the potential not merely of the newly proposed, but also of the earlier developed indicator, and that especially for the benefit of the potential users of indicators, many of whom are not technical experts in the field of indicator development, it is useful to provide a balanced, comprehensive overview of relevant methodological aspects of the various types of indicators, and, most importantly, underline differences in their theoretical assumptions.

In the first part of this paper a framework is presented that allows a systematic comparison of two indicators. This framework is applied in the second part, comparing the original version of the Source Normalized Impact per Paper (SNIP, Moed, 2010, 2011), and a modified version of this metric launched in 2012 (Waltman, Van Eck, Van Leeuwen and Visser, 2013). In addition, it also compares an important component in the two SNIP versions, namely the Raw Impact per Paper (RIP), also indicated as SNIP's numerator, and recently launched by Elsevier's Scopus as a new journal metric, denoted with the acronym IPP (Scopus, 2014). This paper does not aim to give a detailed, technical discussion of the differences between the two SNIP versions. It focuses on main differences in objectives, statistical assumptions and correlations.

It is true that the present writer is the inventor of the original SNIP, and that the new SNIP is to some extent competitive to the old one, claiming to embody substantial improvements. But this does not mean that he is by definition unable to give a balanced view of both types of indicators, and make a serious attempt to identify pros and cons of each type. One may argue that perhaps a more independent author would be more appropriate to compare the two types. This may be true, and the present writer would warmly encourage others to perform such task, and include other types of indicators not explicitly discussed in the current paper. In fact, a systematic comparison of all main types of indicators proposed during the past years would be interesting from a scientific point of view, and most useful for potential users of indicators. But this task is beyond the scope of this paper. Its aim is to illustrate a framework of how such an expanded comparison could be carried out, using a comparison of the original and modified SNIP as an example.

## 2. An Indicator Comparison Report

An Indicator Comparison Report includes 6 sections, listed in Table 1 below. The report's structure reflects the notion that there is no such construct as a 'perfect' indicator, and that one cannot a-priori assume that one indicator is the standard. Hence, it makes sense to systematically compare indicators, and to discuss their differences and similarities in terms of their potential and limitations, and to highlight their theoretical assumptions. Calculating statistical correlations between the measures in specific data samples is an indispensable part of the comparison. One should obtain insight into the extent to which an indicator makes a difference. Next, the report acknowledges that an indicator is described from the point of view of its main objectives, and its application context. This also helps communicating the outcomes of the comparison to users, taking into account their frame of reference. Finally, concept validity – "the degree to which an indicator measure what it claims to measure" – and statistical validity – including concepts such as robustness ("what is the measurement error"), stability ("how stable are the outcomes over time") and consistency ("does an indicator comply with plausible consistency rules") – are conceived as distinct concepts.

**Table 1: Main sections of an Indicator Comparison Report**

| Nr | Section | Main issue |
|---|---|---|
| 1 | Base objective | What is its primary aim? |
| 2 | Concept validity | What does it measure? What are the main theoretical biases? |
| 3 | Application context | How should it be used? |
| 4 | Statistical validity | Key statistical assumptions. What are the main statistical biases? |
| 5 | Target and citing universes | Which sources are included in the database universe? |
| 6 | Degree of correlation | statistical correlation coefficients |

### 3. Comparing the original and the modified SNIP

*3.1 Base concept of the original SNIP*

The base concept of SNIP can be expressed as follows. Many authors have underlined that it is improper to make comparisons between citation counts generated in different research fields, because citation practices can vary significantly from one field to another. For instance, articles in biochemistry often contain over 50 cited references, while a typical mathematical paper has perhaps only 10. This difference is an important factor explaining why biochemical papers are cited so much more often than mathematical ones. Eugene Garfield's view on this is clear. "Evaluation studies using citation data must be very sensitive to all divisions, both subtle and gross, between areas of research; and when they are found, the study must properly compensate for disparities in citation potential (Garfield, 1979, p. 249)".

One of the ways to correct for differences in citation potential across subject fields is the application of fractional citation counting, introduced by Small and Sweeney (1985). The SNIP combines Garfield's notions with the ideas proposed by Small and Sweeney, and also those introduced later by Zitt and Small (2008). A journal's source normalized impact per paper, SNIP, is defined as a ratio of two measures: it divides the number of times a journal's articles are cited – i.e., its raw impact per paper – by the frequency at which papers in the journal' subject field cite other materials – labeled as the subfield's citation potential. In other words, it measures a journal's citation rate per cited reference in documents in the journal's subject field. It must be noted that in the present paper the *term citation potential* relates to the length of reference lists in articles covering a subject field

A journal's subject field is defined as the collection of articles citing the journal. More details about the time window applied in this definition are discussed in subsection 3.4 below ("Statistical Assumptions"). The citation potential in a journal's subject field is a normalized measure calculated in two steps. The first step calculates for each journal the average number of 1-3 year old cited references in the journal's subject field, defined as the collection of articles citing at least one paper published in the journal in the preceeding eight years. In the second step, a normalization is carried out by dividing this measure by its median value across all journals in the database. In this way, the median journal has a (relative) citation potential of 1.0, and the value of its SNIP equals its Raw Impact per Paper. For 50 per cent of journals in the database the SNIP value is higher, and for another 50 per cent it is lower than the Raw Impact per Paper.

*3.2 Differences and similarities between the original and modified SNIP*

The main characteristics, common to both the original and the modified SNIP are:

- They account for differences in citation potential across subject fields and in this way reduce subject field bias of the classical journal impact factor;
- A journal's subject field is defined in a tailor-made manner, taking into account a journal's scope, namely as the total collection of articles citing the journal, and reducing age bias (see Section 3.4);
- They correct for differences in the extent to which the publication database used in the calculations covers a subject field; this dimension was not included in the concept of fractional citation counting developed by Garfield (1979) and Small and Sweeney (1985). Nor is it included in the approach by Leydesdorff and Bornmann (2011) in their study of the effect of fractional counting on journal impact factors. It is partially accounted for by Zitt and Small (2008), as they propose to fractionate on the basis of the number of cited references in JCR-covered journals, a set that is wider than the collection of source journals indexed for the Web of Science.
- They takes into account only peer reviewed articles (normal articles, reviews and conference proceedings articles) published in a journal, both as cited and as citing documents. This feature corrects for the effect of so called "free citations" upon the Thomson Reuters journal impact factor (Moed and Van Leeuwen, 1995).
- The SNIP indicators are calculated at the level of a *journal*, *not* at the level of *individual papers* within a journal. To be specific, a journal's subject field, defined as the collection of articles citing it, is the same for all articles published in it, even when it covers more than one subject field. This illustrates once more that journal citation impact indicators are no good predictors of the citation impact of individual articles in a journal (e.g., Seglen, 1994; Garfield, 1996; Glanzel, 2009).

The current paper does not give a full, detailed description of the two indicators. For details the reader is referred to the articles in which they were introduced. The current paper focuses on differences in base assumptions and their effects upon how the indicators are calculated. Table 2 presents an Indicator Comparison Report for the original and modified SNIP indicator.

**Table 2: Indicator Comparison Table: Original versus Modified SNIP**

| Aspect | Original SNIP | Modified SNIP |
|---|---|---|
| *Objectives and application context* | | |
| Base objective | The indicator shows how classical journal impact factors change when they are corrected for differences in citation potential between their subject fields | Indicator based on source normalization, compliant with intuitively plausible consistency rules and accounting for variability in database coverage over the years |
| Concept validity | Measures journal impact, using the value of the classical journal impact factor as a reference value | Measures relative journal impact, using a 'world average' as reference level |
| Reduction of subject field bias | Subject field normalization is based on the median citation potential across all journals in the database | Subject field normalization is based on the requirement that the (arithmetic) average SNIP across all journals equals 1.0 |
| Application context (intentions) | Shows relativity of classical journal impact factor | Connection to classical journal impact factor is looser; To be used as a subject field-normalized journal impact measure |
| *Statistical assumptions* | | |
| Consistency criterion – 1 | The proposed consistency criterion does not capture the essence of a normalized indicator. Any additional citation carries along information on a journal's subject field, and may even have a negative effect upon the value of the journal impact indicator | Any indicator should comply with the following consistency criterion: If a journal receives an additional citation, the value of its citation impact must increase |
| Consistency criterion – 2 | If two journals merge, a new entity emerges, covering a heterogeneous subject field, and the SNIP value is evaluated relative to that new entity. The outcomes depend upon the underlying configuration | The indicator should also comply with a second consistency criterion: if two journals are merged, the SNIP value of the resulting journal must lie between those of the contributing journals |
| Arithmetic vs. harmonic means | Calculates arithmetic means both for Raw Impact per Paper and for Citation Potential | Calculates arithmetic mean for Raw Impact per Paper but harmonic means for Citation Potential |
| Subject field definition | Defined as the set of articles citing a journal; calculation of citation potential depends on characteristics of individual (citing) articles | Defined as the set of articles citing a journal. But calculation of its citation potential partially depends on characteristics of citing journals rather than individual (citing) articles |
| Reduction of age bias | Subject field is defined as the collection of papers citing 1-8 year old articles published in a journal | Subject field is defined as set of articles citing 1-3 year old articles published in a journal; |
| *Definition of target and citation universes* | | |
| Target universe | All normal articles, reviews and proceedings papers are included | In principle, normal articles, reviews and proceedings papers are included. |

| | | But articles with zero cited references are deleted |
|---|---|---|
| Citing universe analyzed | Citation counts based on all citations recorded in the database | About one third of citing journals is eliminated causing the raw impact per paper to decline with overall 7 % |
| Statistical correlations | | |
| Correlations between the two SNIP versions | The two indicators correlate 0.89-0.97 (Pearson R or Spearman Rho), depending upon the journal set; differences in citing journal sets account for a part of the variance | |

*3.3 Objectives and application context*

A base objective of the original SNIP is to show how values of journal impact factors, and rankings based thereon, change if they are corrected for differences in citation potential, a concept Eugene Garfield used in several of his publications. The SNIP numerator, the Raw Impact per Paper, can be seen as a proxy for the Thomson-Reuters journal impact factor, denoted as *classical journal impact factor* throughout this paper, and SNIP as a corrected journal impact factor. SNIP gives an approximate answer to the question: "What would the value of a given journal's impact factor be if one would take into account that in some fields one cites on average other papers more often than in other domains"?

The modified SNIP is based on two specific indicator *consistency rules* which are further discussed in the next session. Also, it aims to deal with changes over time in the journal coverage of the databases, by eliminating a large number of sources from the citation universe, an aspect which is further considered in Section 3.5 below. The modified SNIP is normalized in such a way that the average SNIP value across the entire database amounts to 1.0. Hence, a value of 1.0 has a specific statistical meaning. In this way, the modified SNIP has an interpretation similar to the relative citation rates based on target or "cited side" normalization: a value above one is above the "world (or database) average". This provides the user with a robust reference level, against which the SNIP value of a particular journal can be compared. For the original SNIP the reference value is the value of the classical journal impact factor. It must be noted that the modified version, as the original one, does not claim to neutralize all possible sources of subject field bias; only the frequency at which researchers in a field cite other papers is taken into account. For instance, SNIP does not take into account the growth of the literature in a subject field (Zitt, 2011).

Informed statements on the application context can be made only on the basis of a thorough investigation of how the indicators are actually being used. Such an investigation goes beyond the scope of the present paper. But one could make an attempt to indicate possible *intentions* of the developers. The original SNIP is designed primarily as a *complement* to the classical journal impact factor rather than being an alternative, and illustrate the latter's relativity. The modified SNIP has perhaps some more pretentions., as its connection to the classical impact factor is looser, and it can be thought to function rather as an independent, alternative journal metric, even though it is based on the same main lines as its original counterpart.

*3.4 Statistical assumptions*

A key assumption underlying the construction of the modified SNIP is that certain *consistency criteria* should hold for an indicator in order to be statistically valid. These criteria are based in common sense or intuition. The *first* consistency criterion that the authors of the modified SNIP apply holds that If a journal receives an additional citation, the value of its impact must increase. Although it does indeed have a certain degree of plausibility, the base notion of the original SNIP would qualify it as a rather simplistic criterion that does not capture the essence of a relative indicator sufficiently well. From the original SNIP perspective, a citation does not come alone. It carries along information about a journal's subject field that has to be taken into account in the calculation of the field's citation potential. When an assessment is made based upon the information available at the moment the citation occurred, the estimated value of the field's citation potential may have increased, and, hence, the SNIP value may have declined. According to the author of the current paper there is nothing counterintuitive in this.

A second consistency criterion applied by Waltman et al. states that if two journals are merged, the SNIP value of the resulting journal must lie between those of the contributing journals. Specifically, if two journals with the same SNIP value are merged, the SNIP value of the resulting journal must be equal to that of its constitutive parts. The authors give an example illustrating that the original SNIP does not in all cases meet this criterion. Two journals with the same SNIP value but differences in citation-per-article rates and citation potentials of a factor 2 merge into a new entity the SNIP of which is about 10 per cent lower than that of the two constitutive journals. The present writer does not see why this second consistency criterion must hold. As the first criterion discussed above, it seems to ignore the complexity of a relative citation measure. When two journals merge, an entirely new entity emerges, covering a heterogeneous subject field, and the SNIP value is evaluated for that new entity. The outcomes depend upon the underlying configuration.

Since these two consistency criteria constitute the rationale for calculating *harmonic* rather than *arithmetic means*, the present writer does not see why harmonic means should have a preferred status. Mingers (2014) rightly observes that "The arithmetic mean depends only on the sum of the values and how many there are. However, the harmonic mean also depends on the spread or dispersion of the values." (Mingers, 2014, p.892). The conclusion seems justified that the added value of harmonic means remains as of yet unclear.

The decision to calculate in the modified SNIP harmonic rather than arithmetic means has important consequences for the way *a journal's subject field is defined*. In order to understand this, the definition of a journal's subject field must be explained first. A journal's subject field is defined as a set of articles citing a particular journal. If one would define this set as the collection of articles citing 1-3 year old other articles in a journal, a strong *age bias* would be introduced in favor of articles citing recent work. Articles citing only 4 years old or older articles in a journal would *not* belong to its subject field.

The original SNIP reduces this age bias by including in a subject field articles citing up to 8 years old articles in a journal. The threshold of 8 years emerges from the pragmatic consideration that it should be more than 3, but not too large, as to avoid the possible disturbing effect of changes in a journal's scope over time, and to reduce negative biases towards recently founded journals. This age bias issue is often ignored in other applications of source-normalized citation measures. For instance, in an interesting article by Zhou and Leydesdorff (2011) applying the notion of source normalization (or fractional counting) to the assessment of scientific institutions rather than journals, this age bias problem is not taken into account.

The implication of the subject definition applied in the calculation of the original SNIP is that a journal's subject field may consist of articles with *no* citations to 1-3 year articles published in the journal. But in this case, harmonic means cannot be calculated.

In the calculation of the original SNIP, the numerator (a journal's raw impact per paper) and denominator (citation potential in a journal's subject field) are as it were calculated separately, and are based on distinct sets of citing publications. An article that does not cite a 1-3 year old paper in a target journal by definition does not contribute to the journal's raw impact per paper, but if this article cites a 4-8 year old paper in the target journal, it does contribute to the calculation of the target journal's subject field.

In the modified SNIP calculation both components are based on the *same* set of citing articles. In fact, Waltman et al. have used another method to reduce the age bias. They multiply each citation with a factor indicating the fraction of 1-3 year old cited references in the *citing* journal. This is a creative solution to the age bias problem, taken from the concept of an audience factor proposed by Zitt and Small (2008). But an important limitation is that, while the base idea of the SNIP is to apply a tailor-made definition of a subject field based on characteristics of individual (citing) *articles* rather than entire *journals*, the weight factors introduced by Waltman et al. do relate to an entire *journal*. The age distribution of citations calculated for the entire journal may be different from that of a subset of citations to a particular target journal.

A final aspect is the robustness of the indicators, particularly their sensitivity to low numbers of "active" cited references in citing journals included in a target journal's subject field. This issue is discussed in the next subsection.

*3.5 Definition of target and citation universes*

The Raw Impact per Paper (RIP) is defined as the average number of citations in a particular citing year to articles, reviews and proceedings papers published in the three previous years. This indicator constitutes the numerator in both the original and the modified SNIP. When Waltman et al. state that the modifications implemented in their new SNIP does not affect the RIP indicator, they mean that their RIP is, like the original SNIP, a simple ratio of sum of citations and number of target documents. But one must be aware that they exclude three sets of journals from the citing universe.

a. Trade journals (262 journals, constituting 1.4 per cent of all 19,816 journals with at least one publication in 2010);
b. Journals that did not publish continuously during four consecutive years (5,568 journals accounting for 28.1 per cent of all journals);

c. Journals for which less than 20 per cent of the publications in the year of analysis have at least one active reference, i.e., a cited reference published in the three preceding years and indexed in Scopus (832 journals, 4.2 per cent).

The elimination of *trade journals* as citing (or target) journals (*set a*) is fully in line with the principle that bibliometric assessment of the contribution to the advancement of scientific-scholarly knowledge should be based on the peer reviewed scientific-scholarly literature, and could be easily implemented in the original SNIP method as well, even though it should be noted that the effect of trade journals is very small. To the best of the present writer's knowledge, in the past few years there are few if any trade journals processed for Scopus

The exclusion of sources which do not publish continuously during 4 consecutive years (*set b*) as *citing* journals is motivated by the need to establish a more "appropriate balance between the number of citing and the number of cited publications" (Waltman et al., 2013, p. 277). The authors show in their mathematical framework that under rather strict assumptions the weighted average value of the modified SNIP of journals in any subject field is approximately 1.0, as outlined above a highly desired feature of the modified SNIP. One of these conditions is that the number of publications made in a subject field is constant over the years. Since journals that do not publish during 4 consecutive years are excluded as target journals, the elimination of this type of sources as citing journals creates a dataset that is more compliant with the assumptions of the model. Assuming that the authors' considerations are captured properly, the present writer feels that the data collection seems to be adjusted to the model rather than the other way around, which from a methodological point of view is questionable. But as outlined in Section 3.3, correcting for changes over the years in the coverage of the database is a crucial objective of the modified SNIP. In Table 5 in Section 3.6, empirical data are presented per subject field on the weighted average values for both the original and the modified SNIP.

The authors of the modified SNIP also argue that some journals (trade journals, magazines and journals with a strong national orientation) may contain a relatively small average number of cited references per publication *(set c)*. When a particular target journal is cited often from this type of journals, the target journal's citation potential is relatively low, and, as a consequence, its SNIP value tends to be high. The authors claim that in this way the SNIP value may benefit more from a citation from a trade journal or an "obscure" national journal, than from a citation from a regular journal.

The present writer agrees that the original SNIP is indeed *sensitive to low numbers of active references* in citing journals. By eliminating these, the indicator becomes more robust, which is indeed an improvement. But one should also realize that a journal with a low citation potential (i.e., a low average number of 1-3 year old cited references published in sources processed for Scopus) is not necessarily "obscure" or non-scientific. For instance, journals in fields that are not well covered by the database may have low citation potentials. But these journals are not necessarily unimportant in their subject fields. It is a very base property of the original SNIP methodology to take this moderate coverage into account, and generate for these journals a citation impact value that is higher than that of the classical impact factor. But the modified SNIP may disregard citation data that are relevant for the assessment of a journal's impact.

Waltman et al. (2013) point towards a limitation of source-normalization as such and of analyzing databases in which large differences in quality and impact between journals exist, and in which a substantial part of journals has hardly any citation impact. This limitation applies to the Scopus database as well. In order to further substantiate this statement, a secondary analysis of two datasets described in the beginning of Section 3.6 was conducted, and percentile values were calculated of the distribution of the total number of received citations among journals covered in Scopus for which SNIP values were calculated and that were active in February 2014. The results are presented in Table 3. The citation counts relate to the SNIP time window, i.e., to citations in a particular year (2010 or 2012) received by articles, reviews and conference papers published in a journal during the three preceding years.

Table 3 shows that 10 percent of the journals received at most 2 citations, and 25 per cent received at most 12, depending upon the year and type of SNIP. These counts are absolute numbers that are *not* corrected for differences in citation potential among subject fields. But the outcomes show that a substantial part of journals is poorly cited. It must be noted that many new journals were added to Scopus in the years before 2010. The citation impact of these journals may not yet have fully matured in 2010 or 2012.

**Table 3: Percentile values of the distribution of citations among journals covered in Scopus**

| *Dataset and year* | *Nr. Journals* | *Percentiles* | | | | |
| --- | --- | --- | --- | --- | --- | --- |
| | | *P10* | *P25* | *P50* | *P75* | *P90* |
| Modified SNIP dataset, 2010 | 17,304 | 1 | 9 | 55 | 255 | 950 |
| Original SNIP dataset, 2010 | 17,769 | 2 | 12 | 64 | 278 | 1,018 |
| Modified SNIP dataset, 2012 | 19,503 | 2 | 11 | 60 | 270 | 987 |

Legend to Table 3: Data relate to journals for which SNIP values were calculated in 2011 and 2013. The citation counts relate to the SNIP time window. P10 indicates the 10th percentile, etc.

*3.6 Statistical correlations*

A quantitative analysis was performed comparing two datasets containing for the (citing) year 2010 original and modified SNIP values, respectively. The SNIP values were available in the Scopus thesaurus of journals for the years 2011 (based on the original SNIP method) and 2013 (based on the modified SNIP approach). The Scopus journal thesaurus is available via http://www.elsevier.com/online-tools/scopus/content-overview.The 2013 Scopus journal thesaurus also contained modified SNIP values for the (citing) year 2010. The data on the SNIP components (number of publications, number of citations) were available at the CWTS website www.journalindicators.com in September 2011 (based on the original SNIP method) and in July 2013 (using the modified approach), respectively.

Tables 4 and 5 present a comparison of SNIP components for the set of journals for which SNIP information is available in both datasets, and which are indicated as "active" in the Scopus thesaurus of journals in February 2014. Table 4 relates to all subject fields combined, while Table 5 presents results per subject field. Since the two SNIP versions applied different methodologies, it does not make sense to compare their values directly. But it is of interest to calculate their weighted mean values across the entire database and compare the outcomes with those presented by Waltman et al. (2013) and see whether they can be *reproduced* (see Table 4). Especially, *differences among subject fields* should be analyzed (Table 5), an aspect the Waltman et al. paper does not provide any information on.

**Table 4: Modified versus original SNIP: Results for all fields combined (n=16,768)**

| *Indicator* | *Modified SNIP method* | *Original SNIP method* | *Difference (%)\** |
|---|---|---|---|
| Total number of articles | 4,259,574 | 4,460,165 | -4.5 % |
| Total number of citations | 8,371,042 | 9,024,382 | -7.2 % |
| Weighted average Raw Impact per Paper (RIP) | 1.97 | 2.02 | -2.9 % |
| Weighted average SNIP | 1.02 | 1.23 | -17.0 % |

*If the value based on the modified SNIP method is denoted as M and that for the original SNIP method as O, Difference is defined as 100*(M-O)/O.

It was found that the total number of publications in all journals is in the modified SNIP calculation conducted in July 2013, 4.5 per cent lower than that obtained for the same journals in the original SNIP. This is due to the fact that the modified SNIP method a priori eliminates in target journals articles with zero cited references (Waltman, 2015). Compared with the original SNIP calculation, the total number of citations included in the modified method is 7.2 per cent lower. This is partially due to the fact that the total number of target articles is 4.5 per cent lower, but also because the modified method discards three subsets of citing journals indicated in Section 3.5.

Focusing on the *weighted raw impact per paper*, the difference between modified and original approach is 2.9 per cent. In this calculation journals are weighted with the number of articles they published. Differences among subject fields are shown in Table 5 below. The differences in Table 4 between the weighted average modified and original SNIP are consistent with findings presented in Waltman et al. (2013). The weighted average modified SNIP is about 1.0, an outcome that is fully intentional, and reflects a key characteristic of this indicator. Its value is 17 per cent lower than the weighted average original SNIP – in other words, the original SNIP value is 21 per cent higher than its modified counterpart, which is a bit lower than the 26 per cent found in Waltman et al.

**Table 5: Original versus modified SNIP: Results per subject field**

| *Subject field* | *Nr. journals* | *Weighted Average Raw Impact per Paper (RIP)* | *Weighted Average SNIP* |
|---|---|---|---|

| | | Modified | Original | DIFF * (%) | Modified | Original | DIFF* (%) |
|---|---|---|---|---|---|---|---|
| *ALL* | 16,768 | 1.97 | 2.02 | -2.9 % | 1.02 | 1.23 | -17.0% |
| Agricultural and Biological Sciences | 1,520 | 1.77 | 1.87 | -5.2 % | 0.97 | 1.13 | -13.9% |
| Arts and Humanities | 1,691 | 0.34 | 0.38 | -8.9 % | 0.65 | 0.58 | 12.3% |
| Biochemistry, Genetics and Molecular Biology | 1,501 | 3.46 | 3.57 | -3.0 % | 1.16 | 1.42 | -17.8% |
| Business, Management and Accounting | 825 | 1.14 | 1.30 | -12.1% | 1.04 | 1.32 | -20.7% |
| Chemical Engineering | 436 | 2.89 | 2.96 | -2.4 % | 1.21 | 1.50 | -19.5% |
| Chemistry | 677 | 2.66 | 2.76 | -3.7 % | 1.07 | 1.34 | -20.5% |
| Computer Science | 1,096 | 0.91 | 1.22 | -25.7% | 0.99 | 1.54 | -36.0% |
| Decision Sciences | 218 | 1.39 | 1.73 | -19.6% | 1.36 | 1.92 | -29.1% |
| Dentistry | 114 | 1.60 | 1.60 | 0.2 % | 1.05 | 1.10 | -4.3% |
| Earth and Planetary Sciences | 831 | 1.84 | 1.91 | -3.7 % | 1.06 | 1.37 | -22.5% |
| Economics, Econometrics, Finance | 644 | 1.10 | 1.25 | -11.7% | 1.15 | 1.43 | -19.7% |
| Energy | 258 | 1.68 | 1.83 | -8.1% | 1.12 | 1.46 | -23.3% |
| Engineering | 1,848 | 1.10 | 1.29 | -14.7% | 0.95 | 1.32 | -28.1% |
| Environmental Science | 871 | 1.98 | 2.10 | -5.8% | 1.09 | 1.29 | -15.7% |
| Health Professions | 297 | 1.51 | 1.56 | -2.7% | 0.94 | 1.07 | -12.9% |
| Immunology and Microbiology | 391 | 3.21 | 3.34 | -3.9% | 1.12 | 1.39 | -19.0% |
| Materials Science | 817 | 1.73 | 1.78 | -2.7% | 0.98 | 1.18 | -17.4% |
| Mathematics | 983 | 0.86 | 0.97 | -11.5% | 0.90 | 1.17 | -22.8% |
| Medicine | 4,569 | 2.37 | 2.39 | -1.0% | 1.05 | 1.13 | -7.3% |
| Multidisciplinary | 82 | 6.53 | 6.34 | 2.9% | 2.03 | 2.72 | -25.4% |
| Neuroscience | 374 | 3.28 | 3.41 | -3.6% | 1.15 | 1.40 | -17.7% |
| Nursing | 442 | 1.59 | 1.56 | 1.6% | 0.85 | 0.86 | -2.0% |
| Pharmacology, Toxicol Pharmaceutics | 557 | 2.38 | 2.48 | -3.9% | 0.93 | 1.02 | -7.9% |
| Physics and Astronomy | 866 | 1.76 | 1.64 | 7.5% | 0.99 | 1.23 | -19.4% |
| Psychology | 850 | 1.85 | 1.97 | -6.3% | 1.11 | 1.34 | -17.8% |
| Social Sciences | 3,414 | 0.82 | 0.92 | -11.1% | 0.93 | 1.00 | -6.7% |
| Veterinary Science | 175 | 1.15 | 1.18 | -2.1% | 0.77 | 0.80 | -3.7% |

*If the value based on the modified SNIP method is denoted as M and that for the original SNIP method as O, DIFF is defined as 100*(M-O)/O.

Table 5 reveals large differences between the original and modified *weighted raw impact per paper*. among subject fields. Differences above 8 per cent are found in arts and humanities, business, management & accounting, computer science, decision sciences, economics, econometrics & finance, energy, engineering, mathematics and social sciences. As indicated in Section 3.5, according to Waltman et al. the largest part of journals removed from the citing universe are those that do not have publications in every year during a four year time span. New sources entering the database during this time period (but not in the first year) belong to this category. Hence, the inclusion of new journals in social sciences and humanities, and of proceedings volumes especially in engineering and computer science would at least partially explain the differences in the amounts of discarded citations among subject fields.

A second explanation relates to the observation that the subject fields showing the largest differences between modified and original total citations are those that are qualified in Moed (2005) as being "not excellently" or "moderately" covered in the Science Citation Index (currently Web of Science). In the written communication in these fields, sources other than journals play an important role, namely preprints (in mathematics), books (in all social sciences and humanities ), and conference proceedings (in the applied and technical fields). This may at least partially explain why the number of active references in Scopus covered journals in these fields is rather low, and hence, why so many of these journals were not considered as sources of citations in the modified SNIP calculation.

Table 5 reveals differences among subject fields, not only in the original weighted average SNIP, but also in the modified version. As outlined in Section 3.1, the SNIP base concept does not claim to neutralize all possible sources of subject field bias; only the frequency at which researchers in a field cite other papers is taken into account. Table 5 enables one to assess to which extent the indicators account for differences between subject fields. Calculating a "normalized" original SNIP by dividing its weighted average in a subject field by the weighted average across all subject fields (1.23), and comparing this normalized original indicator with the weighted average modified SNIP, it is found that the median value of the weighted average modified SNIP across the 27 subject fields is approximately equal to that of the normalized original metric (both are 1.05), but the standard deviation of the former is lower than that of the latter (0.24 versus 0.31). The five subject fields in which the weighted average modified SNIP diverges most strongly from 1.0 are: Multidisciplinary (2.03), Decision Sciences (1.36), Arts and Humanities (0.65), Veterinary (0.77) and Chemical Engineering (1.21). For the original SNIP the five most deviant fields are: Multidisciplinary (2.72), Decision Sciences (1.92), Computer Science (1.54), Chemical Engineering (1.50) and Energy (1.46).

Two conclusions can be drawn. First of all, the modified SNIP indicator is more robust across subject fields than the original SNIP. In this sense, the SNIP modification has been successful. But Table 5 also shows that even the weighted average modified SNIP deviates substantially from 1.0 in several subject fields. This suggests that the assumptions underlying the modified SNIP are only partially valid, even though the citing universe is reduced so strongly. It should be noted that the Scopus subject field classification used in this analysis strongly deviates from the "ideal" configuration assumed by Waltman et al. in the mathematical foundation of their model, namely of being a categorization of articles into non-overlapping fields showing only citation relations within rather than between fields.

**Table 6 Correlations between original and modified SNIP in 5 journal sets**

| *Study* | *Journal set* | *Nr. journals* | *Pearson R* | *Spearman Rho* |
|---|---|---|---|---|
| Waltman et al. | Journals with at least 100 publications during 2007-2009 | 10,331 | 0.93 | n.a. |
| This paper | All journals in study set | 16,768 | 0.90 | 0.91 |
| | Journals with nr. publications during 2007-2009 > 100 | 8,650 | 0.89 | 0.95 |
| | Absolute difference between modified and original nr citations < 10 % | 7,803 | 0.96 | 0.95 |
| | Nr articles>100 and absolute difference between modified and original nr. citations < 10 % | 2,118 | 0.97 | 0.96 |

Table 6 presents Pearson and Spearman coefficients of the correlation between original and modified SNIP in 5 journal sets. Waltman et al. found in a subset of journals with at least 100 publications a Pearson R of 0.93. In the set of journals studied in this paper, Pearson R is slightly lower (0.90). Selecting only journals for which the absolute difference between modified and original nr citations is below 10 per cent, it increases to a value of 0.96 per cent. This table shows first of all that the linear or rank correlations between the original and modified SNIP values are strong. Also, it suggests that the portion of variance in the modified SNIP that is not explained by the original SNIP is to a considerable extent due to differences in underlying citation universes, rather than to the use of harmonic instead of arithmetic means.

### 4. Concluding remarks.

The framework for indicator comparisons proposed and explored in this article provides a comprehensive overview of the main differences and similarities of two indicators. It can easily be expanded to more than two indicators. The comparison shows both the strong points and the limitations of each of the indicators at stake, rather than over-promoting one indicator and ignoring the benefits of alternative constructs. The report focuses on base notions, assumptions, and application modes, which makes it more intelligible to non-experts.

As regards the comparison between the original and the modified SNIP, the conclusion is that none of the two indicators is superior to the other. Both indicators are based on plausible statistical assumptions. The modified SNIP calculation strongly emphasizes the requirement to comply with certain consistency criteria. The author of the current paper has argued above that in his view the proposed criteria do not capture the essence of a relative indicator sufficiently well. Also, the preferred status of calculating harmonic means over arithmetic means in the modified SNIP concept has not been convincingly motivated. The original SNIP can be interpreted as a correction to a subject field bias in the classical, Thomson Reuters journal impact factor. The modified SNIP normalizes citation rates relative to the weighted mean SNIP across all target journals in the database, so that the value of one is a reference value, whereas for the original SNIP the reference value is the value of the classical impact factor.

The original SNIP is based on the total citation volume included in a database (in articles, reviews and conference papers), while its modified counterpart excludes almost one third of journals from the citation universe. On the one hand, in this way the modified SNIP is more robust, and more rigorously accounts for the difficulties of analyzing databases in which large variations exist in annual source coverage, and in citation impact between journals. But on the other hand, differences between weighted average SNIP values between subject fields are still substantial, and relevant citation information is discarded. Both indicators define a journal's subject field as the collection of articles citing a journal, but the calculation of a field's citation potential using the modified approach partially depends on characteristics of citing journals rather than individual (citing) articles.

The original and modified indicators show a strong linear or rank correlation, which, depending on the sample of journals on which the correlation calculation is based, ranges between 0.89 and a value as high as 0.97. In view of this high level of statistical correlation, in practical applications using SNIP as a rough indicator of journal impact, it will not make much difference whether one uses the original or the modified SNIP version, except probably in social sciences & humanities, engineering and computer science, and mathematics, subject fields in which many sources were not included in the calculation of the modified SNIP.

**Acknowledgements**

The author wishes to thank two anonymous referees for their valuable comments on an earlier version of this paper, and also the participants to the Workshop on Bibliometric Indicators, held at Leiden University, and organized by the Center for Science and Technology Studies (CWTS) on September 2, 2014, especially Dr Ludo Waltman, for their feedback.